\documentclass[]{aastex631}

\newcommand{\kms}{\,km\,s$^{-1}$\ }
\newcommand{\HII}{\textsc{Hii}\ }

\begin{document}

\title{Bullet shooting cloud-cloud collision in MIR bubble N65}

\author[0009-0002-8449-1734]{En Chen}
\affiliation{Center for Astrophysics, Guangzhou University, Guangzhou 510006, P. R. China}

\author[0000-0002-5435-925X]{Xi Chen}
\affiliation{Center for Astrophysics, Guangzhou University, Guangzhou 510006, P. R. China}


\author[0000-0001-8060-1321]{Min Fang}
\affiliation{Purple Mountain Observatory $\&$ Key Laboratory of Radio Astronomy, Chinese Academy of Sciences, Nanjing 210034, P. R. China}

\author[0000-0003-3151-8964]{Xuepeng Chen}
\affiliation{Purple Mountain Observatory $\&$ Key Laboratory of Radio Astronomy, Chinese Academy of Sciences, Nanjing 210034, P. R. China}

\author[0000-0002-5077-9599]{Qianru He}
\affiliation{Purple Mountain Observatory $\&$ Key Laboratory of Radio Astronomy, Chinese Academy of Sciences, Nanjing 210034, P. R. China}
\affiliation{School of Astronomy and Space Science, University of Science and technology of China, Hefei 230026, P. R. China}

\author[0000-0003-0559-5982]{Tian Yang}
\affiliation{Center for Astrophysics, Guangzhou University, Guangzhou 510006, P. R. China}

\correspondingauthor{Xi Chen}
\email{chenxi@gzhu.edu.cn}



\begin{abstract}

We report that the formation of the twin-bubble system N65 and N65bis may be caused by the cloud-cloud collision (CCC) from the Bullet Nebula. 
The blue-shifted $^{13}$CO gas component (N65a [47, 55] km s$^{-1}$) is associated with the twin-bubble system, while the red-shifted $^{13}$CO gas component (N65b [55, 62] km s$^{-1}$) is linked to the Bullet Nebula. 
The distinct signatures of CCC, such as the bridge feature, the U-shape cavity and the complementary distribution with displacement, are found between N65a and N65b. 
The collision timescale is estimated to be 1.15 to 2.0 Myr, which is consistent with the dynamical ages of the two \HII regions in N65a (0.73 Myr for N65bis and 1.19 Myr for N65, respectively), indicating their CCC-related origin.     
A total of 354 young stellar objects (YSOs) are founded, which are clustered into eight MST (Minimum Spinning Tree) groups. 
The distribution of M1 (at the post-frontal edge) and M2, M3, M4 (at the pre-frontal edge) suggests that the CCC triggers star formation along the collision path of $b=0\fdg35$, with younger YSOs present at the pre-frontal edge. 
Therefore, the bipolar morphology of the twin-bubble system can be interpreted by the collision of N65a and N65b along $b=0\fdg35$ about 2 Myr ago. 

\end{abstract}



\section{Introduction} \label{sec:intro}

Massive stars ($\geq8$ M$_{\odot}$) are prominent in the Galaxy because of their extreme density, high luminosity, and violent interaction with its surrounding interstellar medium (ISM). 
Despite the importance of massive star formation to the evolution of galaxies, there is currently no certain theory to explain its formation mechanism. 
Recently, both observational and simulation results have indicated that cloud-cloud collision (CCC) is likely to be the main mechanism for massive star formation \citep{Fukui2021}. \cite{Habe1992} proposed a simple CCC model (hereafter the H\&O Model) to explain this mechanism, in which two molecular clouds collide with each other at a supersonic speed and the resulting shocks compress the gas to achieve the ultrahigh density initial condition for massive star formation. Therefore, a typical CCC usually leave behind a single or several O/B-type stars at the collision spot or along the collision path. 

Recent observations have revealed that almost all well-known massive star formation regions (MSFRs) in the Milky Way are likely to link with CCC processes, including M 20 (\citealt{Torii2011}), RCW 120 (\citealt{Torii2015}), M 42 (\citealt{Fukui2018a}), W 43 (\citealt{Kohno2021}), etc. 
The CCC process can rapidly accumulate mass into a small volume, resulting in the formation of ultra-dense massive cores that continue to form massive stars by self-gravity. 
This conclusion is also supported by recent high-resolution three-dimensional (3D) hydrodynamic simulations \citep{Takahira2014, Takahira2018}, which predicted that the dense cores produced by CCC have a power-law mass function. 
Follow-up simulations \citep{Inoue2013, Inoue2018} proposed that the core mass function of dense cores is top-heavy, suggesting that the impact of CCCs lies in the formation of high-mass stars. 
Observationally, the number of O-type stars triggered by CCC is related to a threshold column density \citep{Enokiya2021}. 
An O-type star is formed when the peak column density of a dense core is greater than $10^{22}$ cm$^{-2}$. 
CCC can easily reach this threshold because its supersonic shocks can compress the gas to form a dense layer. 
Therefore, CCC, as an external mechanism, is an efficient factory for massive star formation.

\begin{figure}
	\includegraphics[width=1\textwidth]{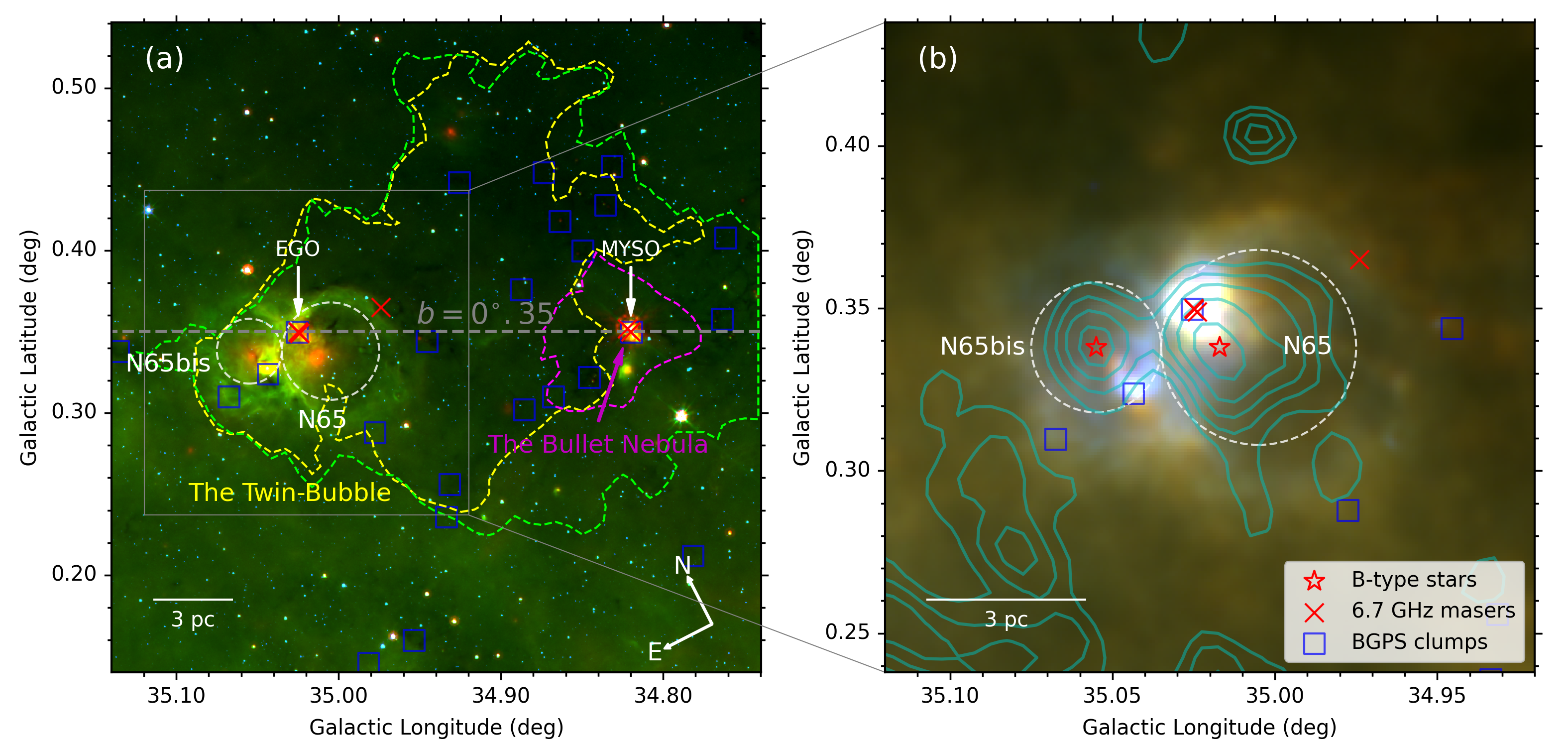}
    \caption{RGB-composite images of the MIR bubble N65. (a) R, G and B colors indicate MIPS 24 $\mu$m, IRAC 8.0 and 3.6 $\mu$m, respectively. The N65 and N65bis bubbles are depicted by white dashed circles. The locations of EGO G35.03+0.35 and MYSO are indicated by white arrows. The yellow and magenta dashed lines show the boundaries (contour at 12 K \kms or $4\sigma$) of the $^{13}$CO gas components that associate with the twin-bubble system and the Bullet Nebula, respectively. And the green dashed line is the total boundary around these two components. (b) Zoom-in region of the twin-bubble system, with its R, G and B colors encoded by $Herschel$ 250, 160, and 70 $\mu$m, respectively. The cyan contours indicate the VGPS radio continuum emission, which start from 17 to 21 K in steps of 1 K. And the open star symbols indicate the peak locations of the VGPS radio continuum emission, which may represent the positions of the exciting B-type stars. To all panels, the blue squares indicate the Bolocam clumps at 1.1 mm, and the red crosses indicate the 6.7 GHz methanol masers extracted from the online tool $MasserDB$. }
    \label{fig:rgb}
\end{figure}

The infrared bubble N65 was initially documented by \cite{Churchwell2006}, which records bubble structures with mid-infrared (MIR) emission in the first ($l=10^{\circ}-65^{\circ},\ |b|\leq1$) and fourth ($l=295^{\circ}-350^{\circ},\ |b|\leq1$) Galactic quadrant of the inner Galactic plane. 
As shown in Figure \ref{fig:rgb}, the N65 bubble is actually a twin-bubble system containing two MIR bubbles, namely N65 and N65bis, located in the first Galactic quadrant with coordinates $(l,b)=(35\fdg005, 0\fdg338$) and $(l,b)=(35\fdg055, 0\fdg338$), respectively. 
An extended green object (EGO G35.03+0.35) reported by \cite{Cyganowski2008} coincident with N65. 
The strong, extended emission in the 4.5 $\mu$m $Spitzer$ IRAC band is usually thought to be produced by shock-excited molecular H$_2$ and CO in protostellar outflows (e.g., \citealt{Smith2006, Davis2007, Chen2010}). 
Hence, an EGO is probably a massive young stellar object (MYSO) driving outflows. 
\cite{Cyganowski2009} reported that the EGO G35.03+0.35 shows a 6.7 GHz Class II maser at the center of the EGO, coincident with a 24 $\mu$m peak, and an UC \HII region, indicating signs of massive star formation. 
\cite{Watson2016} detected an asymmetric non-Gaussian CS ($1-0$) line profile toward the EGO, indicating two unrelated clouds along the line of sight. 
\cite{Deharveng2010} reported that the EGO also coincides with an ATLASGAL 870 $\mu$m continuum peak that lies between N65 and N65bis. 
They interpreted this dust condensation as a sign of compression between the expanding rims of N65 and N65bis.

In addition to the twin-bubble system, our study area also contains a bright IR emission along the same galactic latitude as the EGO at $(l,b)=(34\fdg82, 0\fdg35$), which we call the Bullet Nebula. A bright Red $MSX$ Source (RMS G034.8211+00.3519) is detected toward the Bullet Nebula, which would be considered as a massive young stellar object (MYSO, \citealt{Urquhart2008, Urquhart2011}). 
Although the Bullet Nebula is about 12 pc away from N65, considering their connection in the velocity phase, a CCC process between them is possible. 
According to Phase 3 of the H\&O Model, the two \HII regions in N65 and N65bis appear to be excited by the CCC-triggered O/B-type stars, and their bipolar morphology also provides a clue to the collision path. 
This work will provide evidence for this hypothesis and investigate how the Bullet Nebula collides with N65 and triggers star formation.

The paper consists of five sections. 
Section \ref{sec:data} describes the target and the CO observations and the archival data we used. 
And Section \ref{sec:results} presents the structure and kinematics of molecular clouds, and star formation activities in the study region. 
Section \ref{sec:discussion} discusses the evidence of CCC and its effects on triggering star formation.  
And finally, Section \ref{sec:conclusions} gives the conclusions of this work.

\section{Target and Observations} \label{sec:data}

\subsection{Target}

As mentioned above, our observation target, the MIR bubble N65, consists of two parts: the twin-bubble system and the Bullet Nebula. 
Table \ref{tab:bubbles} summarized the physical properties of the twin-bubble system N65 and N65bis. 
These two bubbles are typical \HII regions with a bipolar morphology along the same galactic latitude of $b=0\fdg35$. 
The bubble structures are fitted with circles according to the VGPS radio continuum emission (see Figure \ref{fig:rgb}(b)). 
This emission traced the ionized gas (inner rim) of the \HII region, with the location of its emission peak representing the possible position of the exciting O/B-type stars. 
According to \cite{Chen2024b}, the spectral type of the exciting star and the dynamical age of the \HII region can be estimated from the integrated flux density and the radius of the radio continuum emission. 
We first calculate the hydrogen-ionizing photon rate $\dot{N}_{LyC}$ from integrated flux density ($S_{\nu}$) of radio continuum emission as follows \citep{Condon1992}, 
\begin{equation}
    \dot{N}_{LyC} \approx 7.54\times10^{46}\ \Big(\frac{T_e}{10^4\ \rm K}\Big)^{-0.45}\Big(\frac{\nu}{\rm GHz}\Big)^{0.1}\Big(\frac{S_{\nu}}{\rm Jy}\Big)\Big(\frac{d}{\rm kpc}\Big)^2\ \rm s^{-1}, 
\end{equation}
where $T_e$ is the electron temperature (typical value of $10^4$ K), $\nu$ is the central frequency of the radio continuum emission (1.4 GHz for VGPS), and $d$ is the distance of the \textsc{Hii} region. 
Compared with Table 1 of \cite{Smith2002}, the spectral types of the exciting stars of N65 and N65bis are B0V and B0.5V, respectively. 
For an evolved \textsc{Hii} region, we estimate its age using a dynamical age from \cite{Spitzer1978}, assuming spherical expansion, 
\begin{equation}
    t_{dyn} \approx 0.057\ \Big(\frac{R_s}{\rm pc}\Big)\Big(\frac{v_i}{\rm 10\ km\ s^{-1}}\Big)^{-1}\left(\Big(\frac{R}{R_s}\Big)^{7/4}-1\right)\ \rm Myr, 
\end{equation}
where $v_i$ is the sound speed of the ionized gas, typically $v_i\approx10$ km s$^{-1}$, $R$ is the radius of the evolved \textsc{Hii} region, and $R_s$ is the radius of the Str$\ddot{\rm o}$mgren sphere, which can be calculated with the formula \citep{Stromgren1939, Dyson1980}, 
\begin{equation}
    R_s = \Big(\frac{3\ \dot{N}_{LyC}}{4\pi\beta_*n_0^2}\Big)^{1/3} \approx 0.74\ \dot{N}_{49}^{1/3}\ n_3^{-2/3}\ \rm pc, 
\end{equation}
where $\beta_*\sim2\times10^{-13}$ cm$^{3}$ s$^{-1}$ is the recombination coefficient for atomic hydrogen. $n_3\equiv [n_0/10^3\ \rm cm^{-3}]$ is the initial number density of gas in a unit of 10$^{3}$ cm$^{-3}$, and $\dot{N}_{49}\equiv [\dot{N}_{LyC}/10^{49}\ \rm s^{-1}]$ is the hydrogen-ionizing photon rate in a unit of $10^{49}\ \rm s^{-1}$. 
Assuming the initial number density of $n_0=5\times10^3$ cm$^{-3}$, the dynamical ages of the \HII regions in N65 and N65bis are estimated to be 1.19 Myr and 0.73 Myr, respectively. 
Notably, the age estimation has a factor of 2 uncertainty due to the assumption of initial density. 
We also calculate the timescale and radius when the \HII regions start to fragment for comparison \citep{Whitworth1994}, 
\begin{equation}
    t_{frag}\approx 1.56\ a_{.2}^{7/11}\ \dot{N}_{49}^{-1/11}\ n_3^{-5/11}\ \rm Myr, 
\end{equation}
\begin{equation}
    R_{frag}\approx 5.8\ a_{.2}^{4/11}\ \dot{N}_{49}^{1/11}\ n_3^{-6/11}\ \rm pc, 
\end{equation}
where $a_{.2}$ is sound speed in a unit of 0.2 \kms. 
Taking the typical sound speed of 0.34 \kms into account, the fragmentation timescale of N65 and N65bis are 1.48 Myr and 1.61 Myr, respectively. 
For N65, $t_{frag} \sim t_{dyn}$ indicates that the N65 \HII region has fragmented, and the substructure of its gas shell (such as EGO) also proves this. 
For N65bis, on the contrary, the \HII region has not yet started to fragment. 
All the calculation results are listed in Table \ref{tab:bubbles}.

The distance of N65 is ambiguous in the literature. 
The \cite{Reid2016} Bayesian distance calculator has given a high likelihood far distance of 10.5 kpc for the source based on its updated Galactic models. 
This distance would give the size of N65 close to 30 pc, significantly larger than the typical bubble size of $<5$ pc reported by \cite{Simpson2012}. 
\cite{Anderson2009} determined the N65 was at the near distance of 3.6 kpc based on the detection of \textsc{Hi} emission/absorption. This distance is consistent with the value of 3.6 kpc used by \cite{Deharveng2010} and 3.4 kpc used by \cite{Brogan2011} and \cite{Cyganowski2011}. 
The parallax measurement of one of the methanol masers in the EGO gives a distance of 2.3 kpc to N65 (\citealt{Reid2016}), consistent with the near distance. 
On the other hand, \cite{Green2011} reported that the kinematic distance of 6.7 GHz methanol masers in the Bullet Nebula is $3.4\pm0.4$ kpc, consistent with the distance of N65. 
Therefore, we adopt the average value of 3.5 kpc as the distance of the twin-bubble system of N65 and the Bullet Nebula in this paper.

\begin{deluxetable*}{cccccccccccc}
\tablecaption{Physical properties of the \HII regions in the twin-bubble system N65 and N65bis. \label{tab:bubbles}} 
\tablehead{
\colhead{Bubble} & \colhead{$l$} & \colhead{$b$} & \colhead{R} & \colhead{$v_{exp}$} & \colhead{S$_{\nu}$} & \colhead{log$(\dot{N}_{LyC})$} & \colhead{Sp.} & \colhead{R$_s$} & \colhead{t$_{dyn}$} & \colhead{R$_{frag}$} & \colhead{t$_{frag}$} \\
\cline{1-12}
\colhead{Name} & \colhead{($^{\circ}$)} & \colhead{($^{\circ}$)} & \colhead{(pc $|$ $\prime\prime$)} & \colhead{(km s$^{-1}$)} & \colhead{(mJy)} & \colhead{(s$^{-1}$)} & \colhead{type} & \colhead{(pc)} & \colhead{(Myr)} & \colhead{(pc)} & \colhead{(Myr)} 
}
\colnumbers
\startdata 
    N65     & 35.005 & 0.338 & 1.83 (108) & 2.8 & 241$\pm$9 & 47.36 & B0V & 0.072 & 1.19 & 2.08 & 1.48 \\
    N65bis  & 35.055 & 0.338 & 1.22 (72)  & 1.8 & 97$\pm$8  & 46.97 & B0.5V & 0.053 & 0.73 & 1.91 & 1.61 \\
\enddata
\tablecomments{(1) Name of the \HII regions in the twin-bubble system. (2)-(3) Galactic coordinates of the \HII regions. (4)-(5) Radius and expanding velocity of the \HII regions. (6) Integrated flux density of the ionized gas at $\nu=1.4$ GHz (or VGPS 21 cm). (7) Hydrogen-ionizing photon rate. (8) Spectral type of exciting O/B-type stars. (9) Radius of the Str$\ddot{\rm o}$mgren spheres. (10) Dynamical age of the \HII regions. (11)-(12) Radius and timescale when \HII regions start to fragment. 
}
\end{deluxetable*}

\subsection{MWISP CO observations}

The observations of CO toward the MIR bubble N65 are part of the Milky Way Imaging Scroll Painting (MWISP\footnote{\url{http://www.radioast.nsdc.cn/mwisp.php/}}) project (see \citealt{Su2019}) conducted by the PMO-13.7m single-dish millimeter telescope at Delingha in China. 
The antenna employed the 3$\times$3-beam Superconducting Spectroscopic Array Receiver (SSAR) system as its frontend \citep{Shan2012} to obtain three CO molecular lines, $^{12}$CO, $^{13}$CO, and C$^{18}$O ($J=1-0$) simultaneously under the on-the-fly (OTF) mode. 
And a total of 18 Fast Fourier Transform (FFT) spectrometers are used as its backend that provide a 1 GHz bandwidth with 16384 channels, producing a frequency interval of 61 kHz and a velocity coverage of $\sim$ 2600 km s$^{-1}$. 
The channel separation and the RMS noise level in a channel width are 0.158 km s$^{-1}$ and 0.48 K for $^{12}$CO data, and 0.166 km s$^{-1}$ and 0.3 K for $^{13}$CO and C$^{18}$O data. 
The telescope has a beam size of 55$^{\prime\prime}$ at 110 GHz ($^{13}$CO and C$^{18}$O) and 52$^{\prime\prime}$ at 115 GHz ($^{12}$CO), with a pointing accuracy of $\sim 5^{\prime\prime}$. 
The final data product was converted into three-dimensional (3D) Flexible Image Transport System (FITS) data cubes of each cell ($30^{\prime}\times30^{\prime}$) with a grid spacing of $30^{\prime\prime}$. 

In this paper, the data cubes are clipped into a field of $24^{\prime}\times24^{\prime}$ centered at ($l, b$) $\sim$ ($34\fdg94$, $0\fdg34$) and a velocity range of $0\sim140$ km s$^{-1}$. 
Throughout this paper, the Galactic coordinate system is utilized and the equinox is J2000.0, and the velocities are all given with respect to the local standard of rest (LSR). 

\subsection{Archival data}

The near-infrared (NIR) J, H and K band photometric data from the UKIRT Infrared Deep Sky Survey (UKIDSS) data release \citep{Lawrence2007} and the mid-infrared (MIR) photometric data from the $Spitzer$ space telescope \citep{Fazio2004} are used to select the disk-bearing young stellar object (YSO) candidates in the survey region. 
The MIR images at IRAC (Infrared Array Camera) 3.6, 4.5, 5.8 and 8.0 $\mu$m and MIPS (Multiband Imaging Photometer) 24 $\mu$m from the $Spitzer$ telescope are used to trace the emission from warm dust. 
And the $Herschel$ images at 70, 160, 250, 350, and 500 $\mu$m from the $Herschel$ infrared Galactic Plane Survey (Hi-GAL) are adopted to trace the emission from cold dust. 
The 1.1 mm radio continuum emission from the Bolocam Galactic Plane Survey (BGPS, \citealt{Rosolowsky2010}) is used to study the dust clumps. 
The angular resolution of the 1.1 mm map is $\sim$33$^{\prime\prime}$. 
The \textsc{Hi} and 21 cm radio continuum emission data (with an angular resolution of $\sim1^{\prime}$) extracted from the VLA Galactic Plane Survey (VGPS, \citealt{Stil2006}) are adopted to trace the ionized gas. 
The 6.7 GHz methanol masers in the survey region were identified using the online tool $MaserDB$\footnote{\url{https://maserdb.net}}, provided by \cite{Ladeyschikov2019}.

\section{Results} \label{sec:results}

\subsection{Structure and kinematics}\label{sec:kin}

\begin{figure}
    \centering
    \includegraphics[width=1\textwidth]{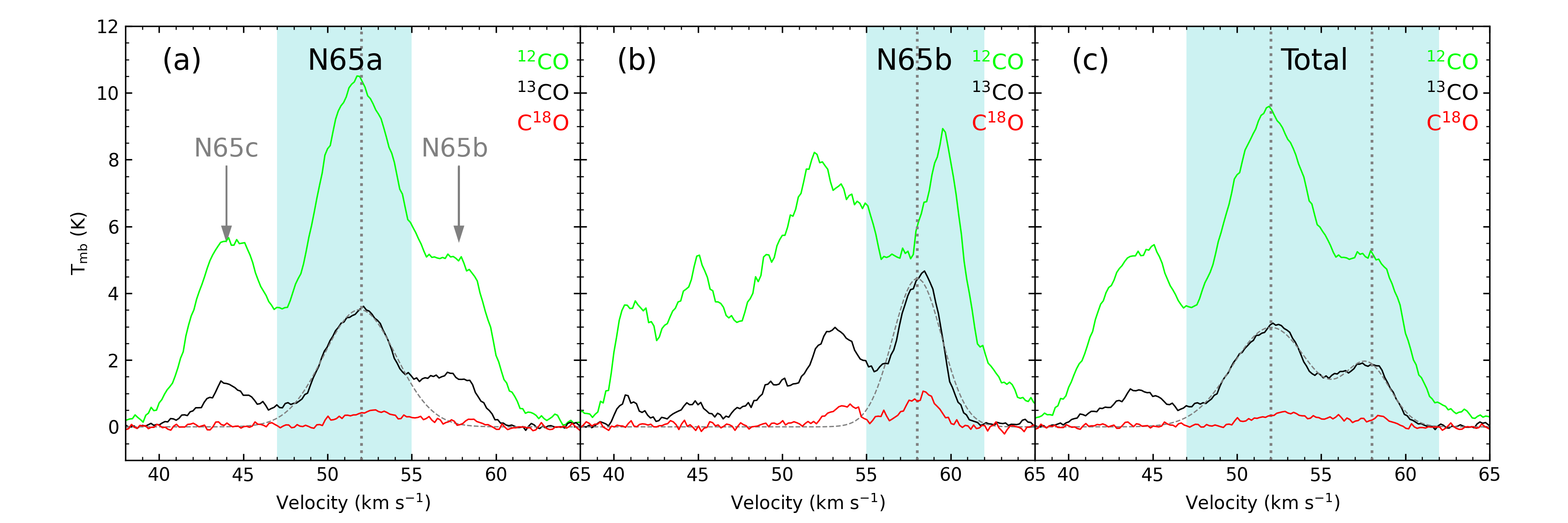}
    \caption{The average spectra of the gas components (N65a, N65b and their combined cloud, respectively) extracted within the boundaries shown in Figure \ref{fig:rgb}(a). To all panels, the green, black and red solid lines indicate the average spectra of $^{12}$CO, $^{13}$CO and C$^{18}$O ($1-0$), respectively. The gray dashed line indicates the gaussian fitting of $^{13}$CO. The two vertical dotted lines are the fitting velocities of N65a (52 \kms) and N65b (58 \kms), respectively. The square shadow highlights the velocity range of each corresponding component. }
    \label{fig:spectra}
\end{figure}

As shown in Figure \ref{fig:rgb}, the study region consists of two parts, namely the twin-bubble system and the Bullet Nebula. 
Figure \ref{fig:spectra} shows the average spectra of the gas components extracted within the boundaries shown in Figure \ref{fig:rgb}(a). 
Three prominent emission peaks are shown in all panels, suggesting the presence of three gas components in the complex with different velocities: N65a [47, 55] \kms, N65b [55, 62] \kms and N65c [39, 47] \kms. 
The system velocities of N65a, N65b and N65c are fitted to be 52, 58 and 44 \kms, respectively.

\setlength{\tabcolsep}{3pt}
\begin{deluxetable*}{cccccccccccccc}
\tablecaption{Physical properties of $^{13}$CO molecular cloud components in the study region. \label{tab:clouds}} 
\tablehead{
\colhead{Cloud} & \colhead{$l$} & \colhead{$b$} & \colhead{$V_{range}$} & \colhead{$V_{LSR}$} & \colhead{$\Delta V$} & \colhead{$\overline{W}_{\rm ^{13}CO}$} & \colhead{$\overline{N}(\rm H_2)$} & \colhead{N$_{pix}$} & \colhead{$A$} & \colhead{$R_{eff}$} & \colhead{$M_{gas}$} & \colhead{M$_{vir}$} & \colhead{$\alpha_{vir}$}\\
\cline{1-14}
\colhead{ID} & \colhead{($^{\circ}$)} & \colhead{($^{\circ}$)} & \colhead{(km s$^{-1}$)} & \colhead{(km s$^{-1}$)} & \colhead{(km s$^{-1}$)} & \colhead{(K km s$^{-1}$)} & \colhead{(10$^{21}$ cm$^{-2}$)} & \colhead{} & \colhead{(pc$^2$)} & \colhead{(pc)} & \colhead{($10^4$ M$_{\odot}$)} & \colhead{($10^4$ M$_{\odot}$)} & \colhead{} } 
\colnumbers
\startdata 
    N65a  & 34.933 & 0.363 & [47, 55] & 52.02 & 5.07 & 17.40 & 9.21 & 709 & 183.7 &  7.65 & 3.79 & 4.11 & 1.08 \\
    N65b  & 34.834 & 0.344 & [55, 62] & 58.02 & 3.41 & 15.85 & 7.83 & 105 & 27.2 & 2.94 & 0.48 & 0.71 & 1.48 \\
    Total & 34.911 & 0.365 & [47, 62] & 53.07 & 9.29 & 23.20 & 11.2 & 1043 & 270.3 & 9.28 & 6.76 & 16.74 & 2.48 \\
\enddata
\tablecomments{(1) Name of the $^{13}$CO gas components. (2)-(3) Galactic coordinates of the gas components, with $l=\Sigma l_i W_i/\Sigma W_i$, and $b=\Sigma b_i W_i/\Sigma W_i$. (4) Velocity range of the gas components. (5)-(6) Gaussian fitting velocity and line width of the average spectra of the gas components. (7)-(8) Average integrated-intensity and column density of the gas components. (9)-(11) Number of covering pixels, area, and effective radius of the gas components, respectively. (12)-(13) Gas mass and virial mass of the gas components. (14) Virial parameters, $\alpha_{vir}=M_{vir}/M_{gas}$. 
}
\end{deluxetable*}

Panel (a) in Figure \ref{fig:spectra} shows the average spectra covering the molecular cloud boundary of N65a (yellow dashed line in Figure \ref{fig:rgb}(a)). 
Within this boundary, the other two peaks (N65b and N65c) are still clearly visible, indicating that they overlap (or partially overlap) each other along the line of sight. 
Panel (b) shows the average spectra within the boundary of N65b, which mainly highlights the peaks at 58 \kms. 
Interestingly, the $^{12}$CO peak shows a notable redshift ($\sim1.5$ \kms) relative to the $^{13}$CO peak, indicating a significant outflow of N65b. 
Panel (c) shows the average spectra within the molecular cloud boundary of N65a and N65b as a whole, whose bimodal structure can be fitted with a double-Gaussian function. 
The physical properties of N65a, N65b and their combined cloud are listed in Table \ref{tab:clouds}. 
Note that since N65c is spatially extended and has no distinct infrared counterpart, it is suspected to be a foreground/background molecular cloud covering the study area along the line of sight, thus, we hereafter omitted further discussion of N65c in the paper.

\begin{figure}
    \includegraphics[width=1\textwidth]{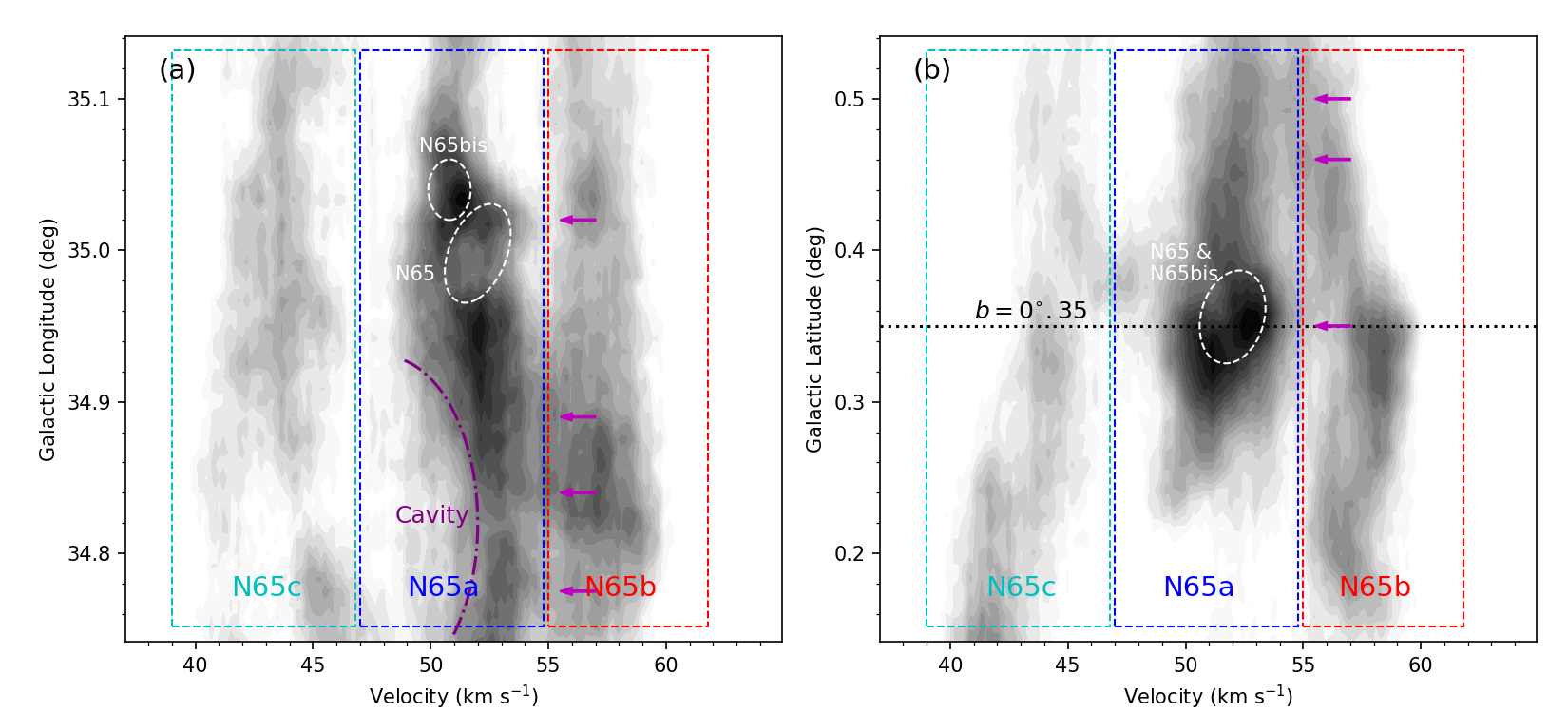} 
    \includegraphics[width=1\textwidth]{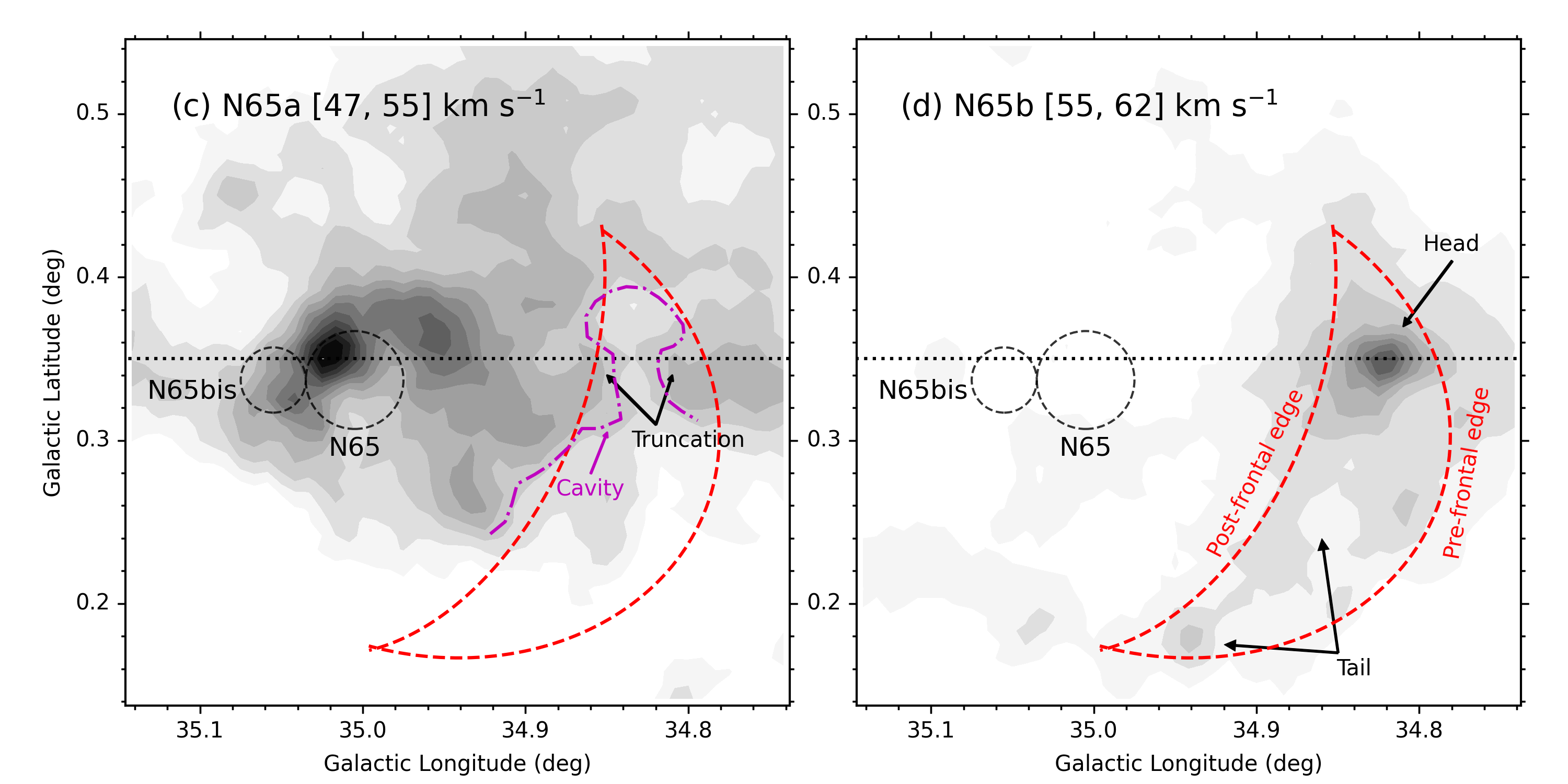} 
    \caption{The kinematic and structure of molecular clouds toward N65. Panels (a) and (b) are $l-$PV and $b-$PV maps of $^{13}$CO gas. The region has three prominent components, namely N65a ([47, 55] km s$^{-1}$), N65b ([55, 62] km s$^{-1}$) and N65c ([39, 47] km s$^{-1}$), depicted by blue, red and cyan dashed squares, respectively. The filled contours are drawn from 15\% to 100\% in steps of 5\% of the maximum mean temperature (3.14 K for $l-$PV map and 3.72 K for $b-$PV map). The white dashed ellipses outline the bubbles of N65 and N65bis. The magenta arrows represent the bridge features connecting N65a and N65b. The purple dash-dotted line highlights the cavity in $l-$PV map. Panels (c) and (d) show the spatial distribution of N65a and N65b, respectively. The filled contours are drawn from 6 (2$\sigma$) to 42 K km s$^{-1}$ in steps of 3 K km s$^{-1}$. The black dashed circles represent the two bubbles. The dotted line represents the galactic latitude of $b=0\fdg35$, which passes through the dense peaks of N65a and N65b simultaneously. The purple dash-dotted line highlights the outline of low density region (cavity with intensity less than 12 K km s$^{-1}$) of N65a. The crescent-like red dotted curves just fit the outline of N65b, which is also drawn in panel (c) for comparison. The component of N65b consists of a dense "Head" and a diffused, elongated "Tail", with the pre-/post-frontal edges along $b=0\fdg35$. }
    \label{fig:PV}
\end{figure}

The longitude-Position-Velocity ($l-$PV) and latitude-Position-Velocity ($b-$PV) maps are shown in Figure \ref{fig:PV}(a) and (b), respectively. 
The corresponding spatial distributions of the $^{13}$CO gas components N65a and N65b are shown in Figure \ref{fig:PV}(c) and (d), respectively. 
In combination with Figure \ref{fig:rgb}(a), the blue-shifted gas component N65a is associated with the twin-bubble system, and the red-shifted gas component N65b is associated with the Bullet Nebula, because of its similarity to infrared emission.

As shown in the $l-$PV and $b-$PV maps, N65a has two distinct ring structures near the coordinates $(l, b)=(35^{\circ}, 0\fdg35)$, which correspond to the twin-bubble system in Figure \ref{fig:rgb}. 
They can be fitted with ellipses whose projection on the velocity axis represents the expanding velocity of the \HII regions. 
It is 2.8 \kms for bubble N65 and 1.8 \kms for bubble N65bis (listed in Table \ref{tab:bubbles}). 
In addition, N65a, as shown in the $l-$PV map, displays a large velocity cavity structure spanning the longitude range of $34\fdg75<l<34\fdg93$, resulting in a velocity gradient of about 0.5 \kms pc$^{-1}$. 
Interestingly, this cavity structure in the velocity phase corresponds to the cavity structure in the spatial phase of the low gas density in N65a (see Figure \ref{fig:PV}(c)), indicating that the gas around this cavity is accelerated due to an external motion, i.e., CCC. 

For N65b, in the $l-$PV map, its gas is mainly concentrated in the cavity range of N65a, with the bridge features connected between them (arrows in Figure \ref{fig:PV}(a)\&(b)), suggesting that the velocity cavity in N65a and its velocity gradient may be caused by the collision from N65b. 
This is the most direct evidence that N65a interacts with N65b. 
The spatial distribution of N65b (see Figure \ref{fig:PV}(d)) exhibits a crescent-like morphology, containing a dense "Head" and a diffused, elongated "Tail". 
The head of N65b is consistent with the Bullet Nebula, with an MYSO embedded. 
The crescent distribution of N65b consists of two edges, the pre-frontal edge and the post-frontal edge, both of which are almost perpendicular to the galactic latitude line of $b=0\fdg35$. 
As shown in Figure \ref{fig:PV}(c), the location and shape of the cavity in N65a is almost identical to N65b, therefore, N65b is complementary to N65a. 
This indicates that N65a and N65b are connected to each other, suggesting the possible CCC process.

\subsection{Young stellar population} \label{sec:YSOs}

Near-/mid-infrared (NIR/MIR) multiband data (including UKIDSS J, H, K bands, and $Spitzer-$IRAC [3.6], [4.5], [5.8], [8.0] four bands, and $Spitzer-$MIPS [24] band) are used to identify young stellar objects (YSOs). 
The identification scheme is based on the work of \cite{Dewangan2018, Baug2018, Chen2024b}, who used four methods to separate YSOs from noise or contamination sources. 
Before running the scheme, we first retrieved these point source catalogs in a $24^{\prime}\times24^{\prime}$ box region centered at $(l, b) = (34\fdg94, 0\fdg34)$. 
And then cross-matched them within $1^{\prime\prime}$ radius to construct a total point source catalog. 

According to the scheme, the first method is to select young embedded protostars using the color-magnitude diagram (CMD) of $[3.6]-[24]/[3.6]$. 
The detailed criteria for this method are given in \cite{Guieu2010} and \cite{Rebull2011}. 
With this method, we obtain 12 Class I, 12 Flat-spectrum, 34 Class II, and 113 Class III sources. 
The second method employs a series of color-color diagrams (CCDs) of four IRAC bands to identify disk-bearing YSOs. 
The criteria are given in the Phase 1 method described in \cite{Gutermuth2009}. 
Consequently, a total of 25 YSOs (13 Class I and 12 Class II) are identified using this scheme. 
The third method is an improvement on the second method, which does not take into account the [8.0] band, but instead constructs the CCD of $[3.6]-[4.5]/[4.5]-[5.8]$ to identify additional YSOs. 
Using the simple criteria \citep{Hartmann2005, Getman2007, Baug2018} of $[4.5]-[5.8]\geq0.7$and $[3.6]-[4.5]\geq0.7$, we obtain a total of 8 Class I YSOs. 
The final method uses the highly sensitive CMD of ${\rm H-K/K}$ to select faint reddened YSOs. 
A cutoff value of ${\rm H-K=3}$ is used for the method, which is estimated from the CMD of the nearby field stars. 
A total of 263 reddened sources meet the criteria of ${\rm H-K>3}$ and are classified as candidates of Class II sources. 

As a result, our YSO identification scheme yields a total of 354 YSOs (33 Class I, 12 Flat-spectrum, and 309 Class II) and 113 Class III in the $24^{\prime}\times24^{\prime}$ survey region. 
The identified Class I, Flat-spectrum, and Class II YSOs are distributed on the $Sptizer$-IRAC 5.8 $\mu$m image and the $^{13}$CO integrated-intensity map in Figure \ref{fig:YSOs_dist}(a) and (b), respectively. 
Note that a remarkable proportion of YSOs are concentrated in subregions.

\subsection{Clustering of YSOs}

To better understand the clustering of YSOs, we performed a surface density distribution analysis and a minimum spanning tree (MST) analysis on the selected sample of 354 YSOs. 
The calculation of the surface density is based on a given number of the nearest-neighbour (NN) YSOs divided by its enclosed circle area. 
More details of the NN technique can be found in \cite{Gutermuth2009}, \cite{Bressert2010} and \cite{Dewangan2018}. 
Taking into account the FOV size ($24^{\prime}\times24^{\prime}$) and distance (3.5 kpc) of the study region, we construct a $5^{\prime\prime}$ grid and a number of 6 NN to calculate the surface density. 
The spatial distribution of the surface density is presented in Figure \ref{fig:YSOs_dist}(a), with its contours drawn at 0.6, 1, 2, 3, 5, 7 YSOs pc$^{-2}$. 
It can be seen that the YSOs are mainly clustered around the twin-bubble system and the Bullet Nebula. 
Interestingly, the YSOs around the N65 bubble are distributed in a semi-ring-like structure with an opening toward the Bullet Nebula, suggesting that the formation of these YSOs may be induced by the expansion of the N65 \HII region. 
However, several YSO groups are distributed around the Bullet Nebula with remarkable offsets, suggesting that the formation of these YSO groups is unlikely to be caused by the self-collapse of the Bullet Nebula, but may be triggered by other external mechanism, i.e., CCC.

MST technique is beneficial for extracting structures with chain-like or filamentary distribution. 
Based on \cite{Chen2024}, we set a cutoff distance of $67^{\prime\prime}$ (corresponding to 1.14 pc) and a minimum number of group members of 10 to perform MST. 
As a result, 8 MST groups were extracted from the selected YSOs. 
Their physical properties are listed in Table \ref{tab:MST}, and their spatial distributions are shown in Figure \ref{fig:YSOs_dist}(b). 
The distribution of these MST groups is consistent with that of surface density, but is better in tracing the coherent clumpy structure. 
For example, the chain-like M1 is consistent with the gas distribution of N65b, which better reflects the formation correlation between them, although it consists of several independent surface density groups. 
The relationship of these MST groups to the CCC will be discussed in Section \ref{sec:SF}.

\section{Discussion} \label{sec:discussion}

\subsection{Signatures of CCC}

Many studies (\citealt{Torii2015, Torii2017, Fukui2018b, Fukui2021, Chen2024, Chen2024b}) have pointed out that the velocity bridge feature, the spatial complementary distribution with displacement, and the U-shape cavity structure are powerful indicators of cloud-cloud collision (CCC) between two individual molecular clouds. 
Based on this argument, we match these signs of collision by investigating the kinematic structure of the molecular cloud and the spatial distribution of each components.

\subsubsection{Broad bridge feature}\label{sec:bridge}

The broad bridge feature is the molecular gas with intermediate velocity that slows down during collision. 
The kinetic motion of two colliding molecular clouds will transform into a turbulent motion, increasing the velocity dispersion near the colliding interface (\citealt{Fukui2018b, Chen2024}). 
Broad bridge feature will have a finite lifetime due to the stellar feedback, the conversion of gas into stars, and the duration of collision. 
\cite{Haworth2015a} proposed that the broad bridge feature is resilient to the effects of radiative feedback, at least to around 2.5 Myr after the formation of the first massive (ionizing) star can cause a significant impact. 
Their follow-up simulation \citep{Haworth2015b} shown that the lifetime of the broad bridge is more likely to be determined by the lifetime of the collision rather than the radiative feedback disruption timescale. 
Since the N65 \HII region has a dynamical age of 1.19 Myr, stellar feedback has not yet destroyed the bridge feature, so the broad bridge between N65a and N65b may originate from a CCC process. 

\begin{figure}[hb]
    \centering
    \includegraphics[width=1\textwidth]{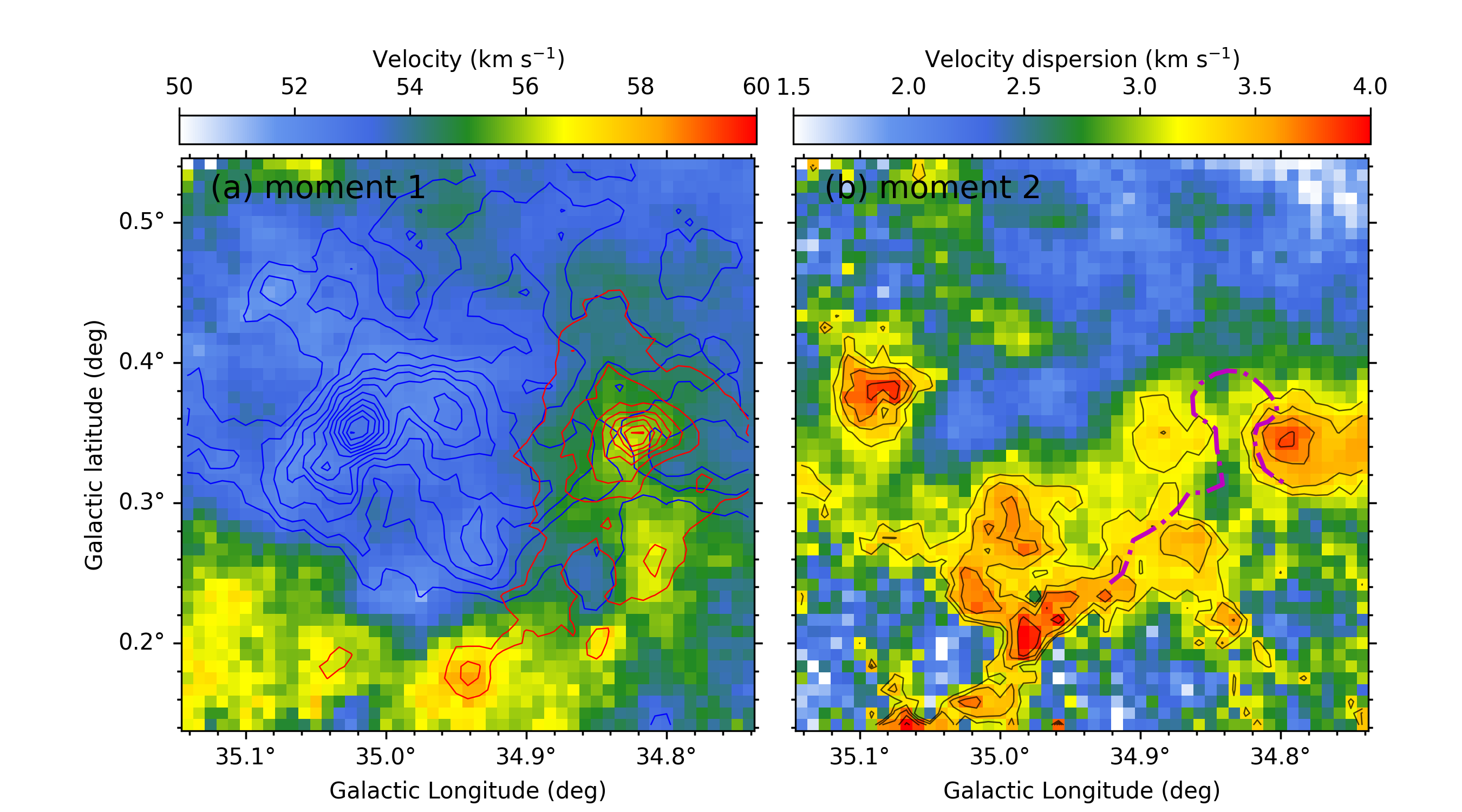}
    \caption{Moment maps of the combined cloud of N65a and N65b. Panel (a) shows the moment 1 map. The blue and red contours indicate the integrated-intensity maps of N65a and N65b, respectively. The contour levels start from 9 to 42 K \kms in steps of 3 K \kms. Panel (b) shows the moment 2 map. The contours indicate the gas with large velocity dispersion, whose levels start from 3.2 to 4 \kms in steps of 0.2 \kms. The dotted-dashed line indicate the boundary of the U-shape cavity of N65a. }
    \label{fig:moments}
\end{figure}

The broad bridge features are shown by the arrows in Figure \ref{fig:PV}(a)\&(b) that connect the two individual components of N65a and N65b. 
Interestingly, this connection is evident across the longitude range ($34\fdg75<l<34\fdg93$) of the velocity cavity of N65a, suggesting that it is likely caused by the collision of N65b. 
The increased velocity dispersion along the U-shape cavity of N65a or the boundary of N65b (see the moment 2 map in Figure \ref{fig:moments}(b)), suggests the possible colliding interface of them. 
Furthermore, the gas with high velocity dispersion ($>3.5$ \kms, contours in Figure \ref{fig:moments}(b)) is mainly distributed along the galactic latitude and is highly correlated with the shape of N65b, suggesting that N65b moves along the galactic latitude. 
N65b collides with N65a along its motion path, thereby increasing the velocity dispersion on it. 
This moving track along the collision path is also observed in the distribution of YSOs, which will be discussed in Section \ref{sec:SF}.

The fact that the two clouds appear in the moment 1 map simultaneously (see Figure \ref{fig:moments}(a)) and are distributed along the galactic latitude indicates that their collision direction is also along the galactic latitude, with the collision angle ($\theta$) almost perpendicular to the line of sight ($45^{\circ}\sim90^{\circ}$). 
The morphologies of the two circular bubbles in N65a (see Figure \ref{fig:rgb}) also support this almost face-on projection. 
According to \cite{Fukui2021}, the maximum collision speed ($v$) should be less than 30 \kms, thus the collision angle (${\rm cos}\theta=v_{||}/v$) between the two clouds should be less than $80^{\circ}$. 
Considering that the relative velocity of the two clouds along the line of sight is only 6 \kms ($v_{||}$), to be safe, their collision angle should be less than $60^{\circ}$, hence the collision angle can be limited to $45^{\circ}\sim60^{\circ}$. 
As a result, the actual collision speed ($v=v_{||}/{\rm cos}\theta$) of the two clouds is between 8.5 and 12 \kms, indicating that N65b probably collided with N65a at a typical CCC collision velocity of 10 \kms. 

\subsubsection{Complementary distribution and U-shape cavity}

As shown in Figure \ref{fig:PV} (a) and (c), the velocity cavity of N65a corresponds to the U-shape cavity structure in its spatial distribution.  
Coincidentally, the crescent-like distribution of N65b (see Figure \ref{fig:PV}(d)) complements this cavity suggesting a physical connection between the two clouds. 
The spatial complementary distribution of two individual components is another evidence of CCC (\citealt{Fukui2018b, Takahira2018}). 
Therefore, collision from N65b not only creates a U-shape cavity (similar in shape and size to N65b) in N65a, but also increases the local velocity of N65a at the corresponding position, with a velocity gradient of about $0.5$ km s$^{-1}$pc$^{-1}$. 
As shown in Figure \ref{fig:PV}(a), the U-shape cavity nearly truncates N65a into two parts. 
The truncation position corresponds to the densest clump (\textquotedblleft Head\textquotedblright) of N65b, indicating the main collision spot of this position. 
This is also supported by the large velocity dispersion on both sides of the truncation position in the moment 2 map in Figure \ref{fig:moments}(b). 
Comparing the relative positions of N65a and N65b, it can be seen that N65b and U-shape cavity are located in the low-density region of N65a, or the diffused tail of N65a, so the collision between N65b and N65a may have occurred much earlier. 
The current collision would be that N65b sweeps across N65a along the galactic latitude and continues to collide with its tail. 
This explains the bullet-like clumpy structure (\textquotedblleft Head\textquotedblright) of the dense gas in N65b and the crescent-like distribution of its diffuse gas (\textquotedblleft Tail\textquotedblright). 
This pre-frontal (the bullet clump) and post-frontal (the diffuse gas) structure is probably caused by the bow shocks of the collision. 
The gas in front of the shock wave is compressed to form a high-density region, while the gas behind is diffused to form a low-density region.

We can further estimate the dynamic state (fragment or collapse) of molecular clouds by investigating their virial parameters. 
According to \cite{Bertoldi1992}, the virial parameter is defined as the ratio of twice the kinetic energy to the gravitational energy of a cloud, 
\begin{equation}
    \alpha_{vir}\equiv\frac{2E_{kin}}{E_{grav}}=\frac{5\delta_{v}^{2}R}{GM}, 
\end{equation}
where $E_{kin}$ and $E_{grav}$ are the kinetic and gravitational energy of the molecular system, respectively. And $\delta_{v}$ is the velocity dispersion, $R$ is the effective radius of the molecular cloud, $M$ is the total mass of the cloud, and $G$ is the gravitational constant. 
According to \cite{Kauffmann2013}, $\alpha_{vir}=1$ is gravitationally in virial equilibrium and $\alpha_{vir}\approx2$ is marginally gravitationally bound, while $\alpha_{vir}\gtrsim2$ indicates an unbound cloud. 
The virial parameter can also be written as $\alpha_{vir}=M_{vir}/M$ (\citealt{Wilson2009}), by defining a virial mass ($M_{vir}$) as, 
\begin{equation}
    M_{vir} = \frac{5\delta_{v}^{2}R}{G}\approx210\ \Big(\frac{R}{\rm pc}\Big)\ \Big(\frac{\Delta V}{\rm km\ s^{-1}}\Big)^2\ {\rm M_{\odot}}, 
\end{equation}
where $\Delta V=\sqrt{8ln2}\ \delta_v$ is the average line width (FWHM) of the cloud. The effective radius $R$ can be estimated through the enclosed area ($A$) of the cloud, $R=\sqrt{A/\pi}$. 
We obtain the cloud area $A$ by counting the number of pixels surrounded by the closed contour at the integrated-intensity of 12 K \kms ($4\sigma$) for each cloud (N65a, N65b and together as a whole cloud). 
The boundaries of these three molecular clouds are shown in Figure \ref{fig:rgb}(a), depicted by yellow, magenta, and green dashed lines, respectively. 
The gas mass of each cloud is also calculated within these boundaries by assuming local thermodynamic equilibrium (LTE), 
\begin{equation}
    M_{gas}=\mu m_{\rm H} d^2 \int N(\rm H_2) d\Omega, 
\end{equation}
where $\mu=2.8$ is the mean molecular weight per hydrogen molecule, $m_{\rm H}$ is the mass of the atomic hydrogen, $d$ is the distance of molecular cloud and ${\rm d\Omega}$ is the solid angle element. $N(\rm H_2)$ is the column density of the molecular hydrogen, which can be calculated by the following equation \citep{Wilson2009}, 
\begin{equation}
    N({\rm H_2}) = 2.42\times10^{14}\ R_{hc}\ \frac{1+0.88/T_{ex}}{1-{\rm exp}(-5.29/T_{ex})}\ \int T_{mb}{(\rm ^{13}CO)d}v , 
\end{equation}
where $T_{ex}$ is the excitation temperature which is estimated from the peak temperature of the optical thick $^{12}$CO emission, $T_{ex}=5.53/ln(1+5.53/(T_{peak}+0.819))$, $T_{mb}(\rm ^{13}CO)$ is the main beam temperature of the $^{13}$CO spectra, ${\rm d}v$ is the velocity element and $R_{hc}$ is the abundance ratio of [${\rm H_2}$/$^{13}$CO], which is estimated to be $6.1\times10^5$, according to the assumptions of a constant [${\rm H_2}$/$^{12}$CO] abundance ratio of $1.1\times10^4$ \citep{Frerking1982} and a [$^{12}$C/$^{13}$C] isotopic ratio gradient in our Galaxy \citep{Milam2005}, [$^{12}$C/$^{13}$C] $=6.21D_{GC}+18.71$. 
$D_{GC}$ is the distance to the Galactic center of the object, for N65, $D_{GC}=5.83$ kpc. 
From the above formulas, we can calculate the column density and gas mass of each molecular cloud component in N65. 
And as a result, we list the physical properties of these three clouds in Talbe \ref{tab:clouds}. 

The virial parameters of N65a and N65b are 1.08 and 1.48, respectively, which are less than 2, indicating that they are gravitationally bound as individual clouds. 
At this point, the kinetic energy of molecular clouds is not enough to support against self-gravity, leading to instability: they will collapse to form stars. 
The presence of MYSOs in both clouds also supports this. 
In contrast, when considering the two clouds as a whole, the virial parameter is 2.48, indicating that the complex is gravitationally unbound. 
This leads to another instability: it will fragment and expand. 
This is consistent with the last phase of a typical CCC process (\citealt{Fukui2021}) in which the two clouds separate from each other.

\subsubsection{Displacement along \texorpdfstring{$b=0\fdg35$}{b=0.35}}

\begin{figure}[hb]
	\includegraphics[width=0.95\linewidth]{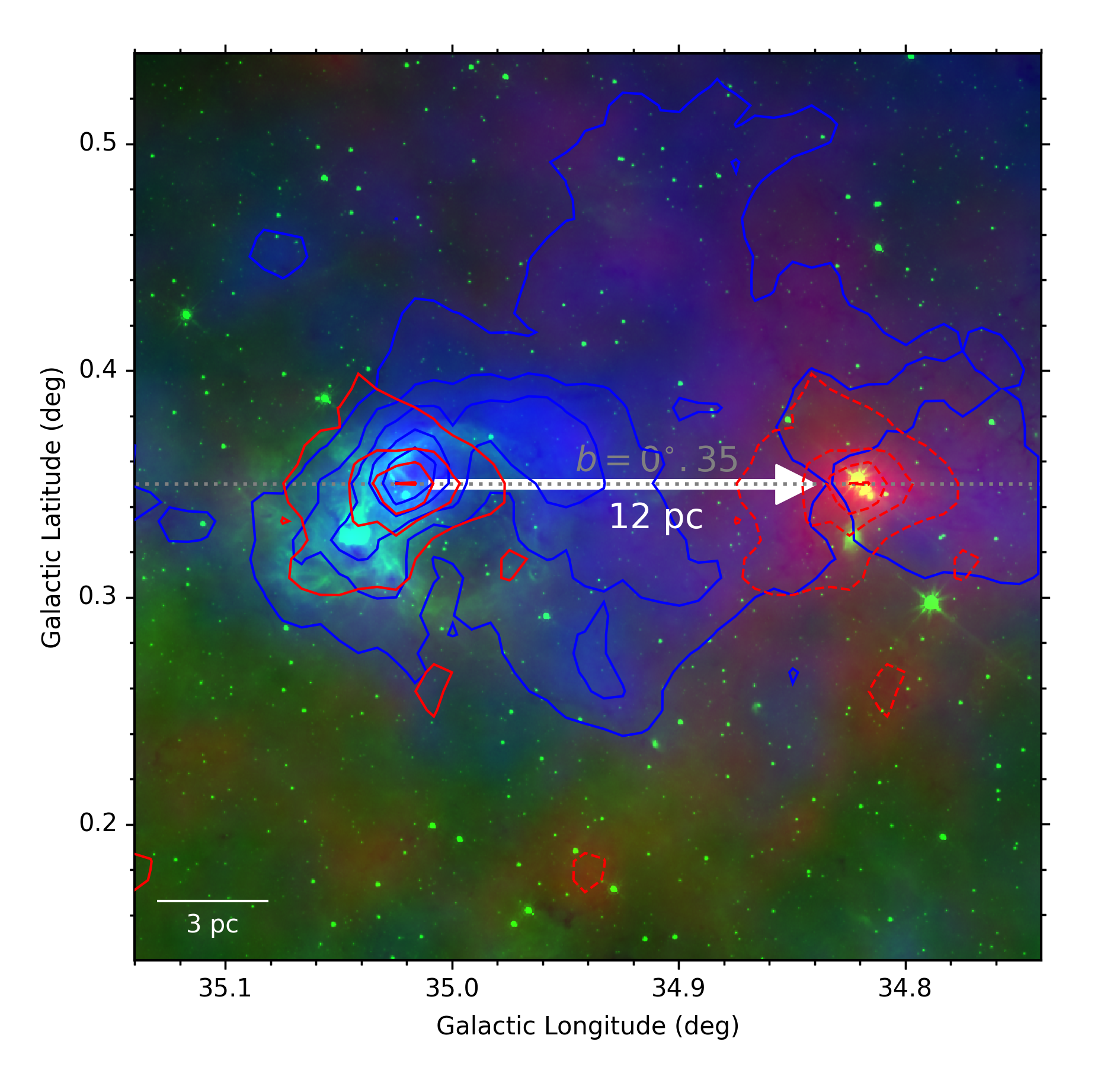}
    \caption{Displacement of N65b along $b=0\fdg35$. The background image is the RGB-composite map with its R, G and B colors encoded by N65b, $Spitzer-$IRAC 8.0 $\mu$m and N65a, respectively. The solid blue contours indicate the $^{13}$CO integrated-intensity map of N65a, while the dashed red contours indicate that of N65b. The solid red contours is the displacement of N65b along the galactic latitude of $b=0\fdg35$ (the dotted line), with a length of 12 pc (the white arrow). The contour levels in the figure are drawn from 12 (4$\sigma$) to 42 K km s$^{-1}$ in steps of 6 K km s$^{-1}$. }
    \label{fig:disp}
\end{figure}

As mentioned in Section \ref{sec:bridge}, N65b collides with N65a along the galactic latitude. Interestingly, the main axis of the two bubbles (N65 and N65bis) and the densest peaks of N65a and N65b are almost on (or parallel to) the same galactic latitude line of $b=0\fdg35$ (see Figure \ref{fig:PV}(c)\&(d)). 
Considering that the collision direction of the CCC remains in the same direction without interference, it is possible that N65a and N65b collided along the galactic latitude of $b=0\fdg35$. 
Based on this argument, we can easily interpret the bipolar morphology of the twin-bubble system and the distribution of dense clumps along the $b=0\fdg35$ galactic latitude line. 
Figure \ref{fig:disp} shows the displacement of N65b along the opposite direction of the collision, which represents the footprint of N65b when the collision started. 
The displacement length is $L=12$ pc. 
Considering that the velocity separation of the two clouds along the line of sight is $v_{||}=6$ \kms, and the collision angle of incidence is $\theta=45^{\circ}\sim60^{\circ}$, the time when the collision starts can be estimated as $t=L/(v_{||}\ {\rm tan}\theta)\approx1.15\sim2.0$ Myr. 
This is consistent with the dynamical ages of the \HII regions for N65 (1.19 Myr) and N65bis (0.73 Myr) in N65a, suggesting that their formation may be induced by CCC.

\subsection{Triggered star formation} \label{sec:SF}

\begin{figure}[hb]
	\includegraphics[width=1\textwidth]{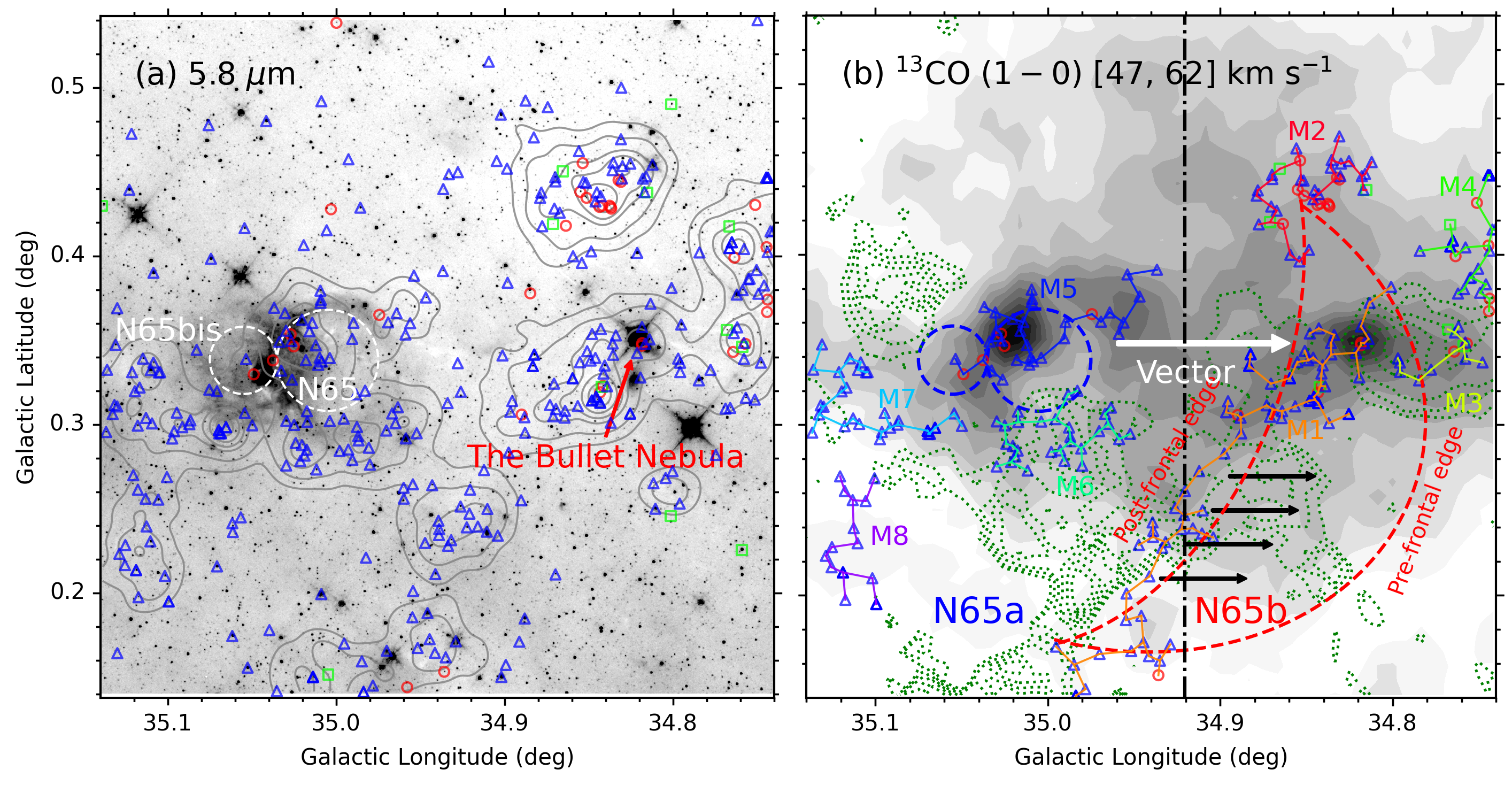}
    \caption{Spatial distribution of YSOs in the study region. Panel (a) shows the spatial distribution of YSOs overlapped on the 5.8 $\mu$m image. Class I, Flat-spectrum and Class II YSOs are marked by red circles, green squares and blue triangles, respectively. The gray contours indicate the 6 nearest-neighbors surface density map of the YSOs, drawn at 0.6, 1, 2, 3, 5, and 7 YSOs pc$^{-2}$. Panel (b) shows the spatial distribution of the YSO MST groups overlapped on the $^{13}$CO integrated-intensity map integrated from 47 to 62 km s$^{-1}$. The dotted contours indicate the gas with large velocity dispersion from Figure \ref{fig:moments}(b). The dashed-dotted line simply represents the colliding interface of N65a and N65b, with the colliding vector depicted by the white arrow. The black arrows represent the possible offsets of M1 caused by the dragging effect of the collision. The pre-/post-frontal edges of N65b are depicted by the red dashed lines. }
    \label{fig:YSOs_dist}
\end{figure}

We have conducted a census of YSOs in the study area in Section \ref{sec:YSOs}. 
A total of 354 YSOs were found, of which 33 are Class I, 12 are Flat spectrum, and 309 are Class II sources. 
According to MST analysis, these YSOs are basically clustered into 8 MST groups. 
The physical properties of the YSO MST groups are listed in Table \ref{tab:MST}. 
As shown in Figure \ref{fig:YSOs_dist}(b), M1 is distributed in an elongated chain shape, and its extension direction is basically the same as that of N65b, with a slight offset (post-frontal edge, black arrows in the figure) along the galactic latitude direction. 
On the other hand, M2, M3, and M4 are located at the head of the N65b (pre-frontal edge). 
This pre-/post-frontal distribution suggests that N65b may be moving along the galactic latitude. 
Moreover, the distribution of M1 and M3 is consistent with that of gas with high velocity dispersion, suggesting that they are affected by the CCC. 
Therefore, it can be inferred that N65b collided with N65a along the galactic latitude direction and triggered the formation of M1 and M3 along the collision path.

\begin{deluxetable*}{ccclccccc}
\tablecaption{Physical properties of the YSO MST groups. \label{tab:MST}} 
\tablehead{
\colhead{MST} & \colhead{$l$} & \colhead{$b$} & \colhead{R$_{eff}$} & \colhead{$\rm N_I$} & \colhead{$\rm N_{II}$} & \colhead{$\rm N_{YSO}$} & \colhead{I/II} & \colhead{$\overline{\alpha}$} \\
\cline{1-9}
\colhead{ID} & \colhead{($^{\circ}$)} & \colhead{($^{\circ}$)} & \colhead{(pc $|$ $\prime\prime$)} & \colhead{} & \colhead{} & \colhead{} & \colhead{} & \colhead{} 
}
\colnumbers
\startdata 
    M1  & 34.884 & 0.282 & 4.31 (254) & 7$\pm$3 & 77$\pm$9 & 84$\pm$9 & 0.09$\pm$0.04 & -0.35\\
    M2  & 34.846 & 0.437 & 1.88 (111) & 14$\pm$4 & 26$\pm$5 & 40$\pm$6 & 0.54$\pm$0.19  & -0.12\\
    M3  & 34.771 & 0.342 & 1.02 (60) & 4$\pm$2 & 10$\pm$3 & 14$\pm$4 & 0.40$\pm$0.23  & -0.18\\
    M4  & 34.754 & 0.405 & 1.49 (88) & 6$\pm$2 & 20$\pm$4 & 26$\pm$5 & 0.30$\pm$0.12 & -0.29\\
    M5  & 34.996 & 0.355 & 2.19 (129) & 5$\pm$2 & 29$\pm$5 & 34$\pm$6 & 0.17$\pm$0.08 & -0.40\\
    M6  & 35.006 & 0.294 & 1.73 (102) & 0 & 29$\pm$5 & 29$\pm$5 & 0 & -0.72\\
    M7  & 35.099 & 0.310 & 1.83 (108) & 0 & 32$\pm$6 & 32$\pm$6 & 0 & -0.65\\
    M8  & 35.113 & 0.230 & 1.44 (85)  & 0 & 16$\pm$4 & 16$\pm$4 & 0 & -0.63\\
\enddata
\tablecomments{(1) ID of the YSO MST groups. (2)-(3) Average coordinates of the MST groups. (4) Effective radius of the MST groups. (5)-(7) Number of members of the MST groups in terms of Class I (including Flat-spectrum sources), Class II and total, respectively. The errors are estimated from the Poisson error, $\sigma=\sqrt{N}$. (8) Ratio of the number of Class I to Class II in the MST groups. (9) Average spectral indices of the MST groups. 
}
\end{deluxetable*}

\begin{figure}[hb]
    \centering
    \includegraphics[width=0.48\linewidth]{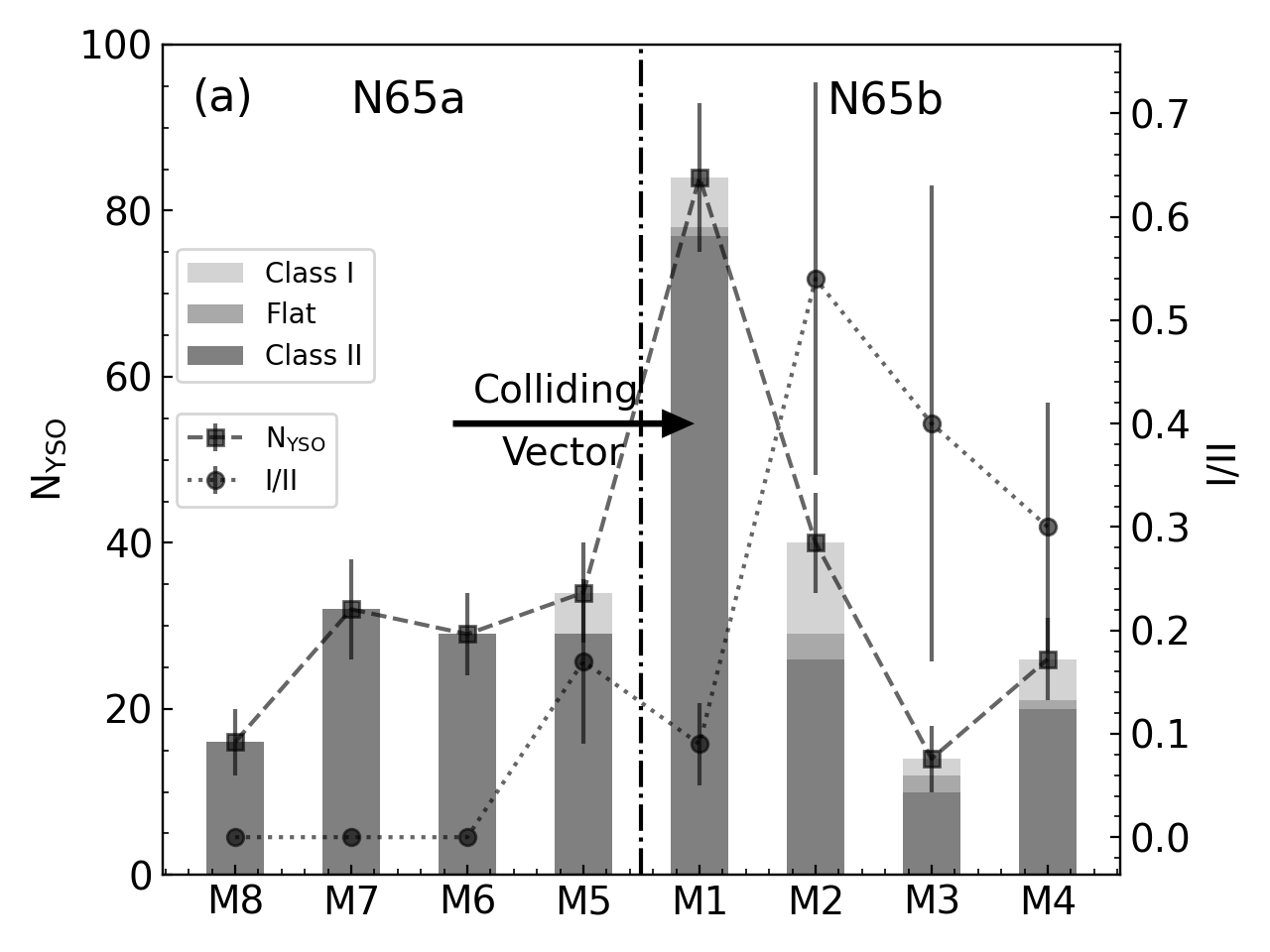}
    \includegraphics[width=0.48\linewidth]{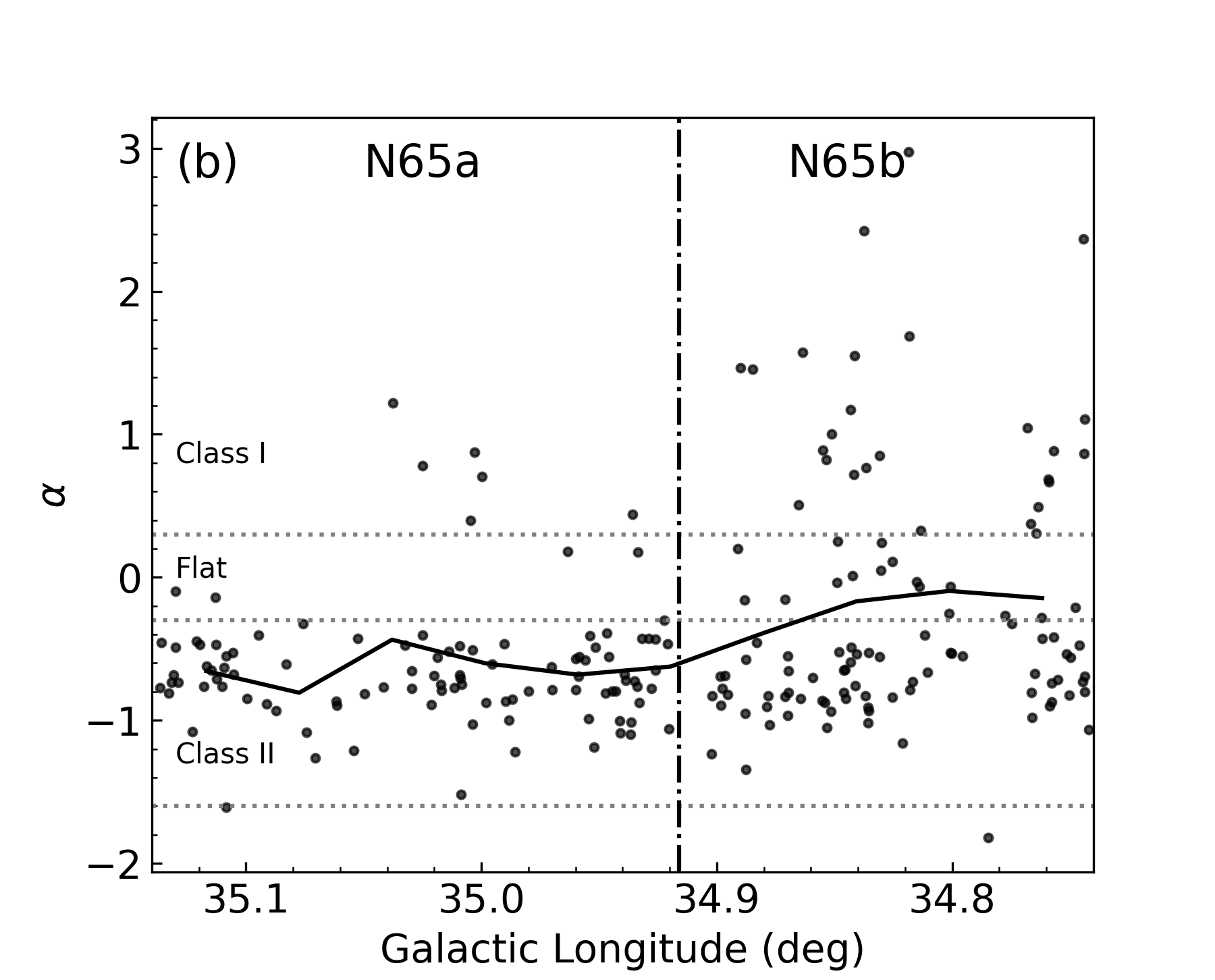} 
    \caption{Star formation along the collision path (or the galactic longitude). Panel (a) shows the distributions of the MST groups along the collision path in terms of the number of group members (the dashed line) and the ratio of number of Class I to Class II sources (the dotted line). The bar chart shows the number of YSOs classifications (Class I, Flat, and Class II) for each MST group. Panel (b) shows the distribution of $\alpha$ indices along the galactic longitude. The dotted lines separate the YSOs into four classes, with $\alpha>0.3$ as Class I, $0.3>\alpha\geq-0.3$ as Flat-spectrum, $-0.3>\alpha\geq-1.6$ as Class II, and $\alpha<-1.6$ as Class III YSOs. The solid line indicate the average $\alpha$ indices along the galactic longitude. To all panels, the vertical dotted-dashed line represents the colliding interface, which divides the study region into two parts, the N65a-dominated and N65b-dominated regions. }
    \label{fig:trend}
\end{figure}

In case of N65a, both M5 and M6 are distributed around the edges of the bubble N65, indicating that their formation is influenced by the expansion of the N65 \HII region. 
In addition, the location of M6 is consistent with that of the gas with high velocity dispersion, indicating that the formation of M6 may also be affected by CCC in addition to the expanding \HII region. 
However, M7 and M8 are distant from the twin-bubble system, suggesting that their formation may be less affected by the two \HII regions. 
Since M6, M7 and M8 only contain relatively old Class II YSOs, with an average age of $2\pm1$ Myr (\citealt{Evans2009}) consistent with the aforementioned collision timescale of about 2 Myr, it indicates that M6, M7 and M8 may be the leftover YSOs from the collision path. 

Interestingly, almost all signs of massive star formation in N65, including two B-type stars, two maser sources, and one MYSO, are distributed along the galactic latitude of $b=0\fdg35$, suggesting that massive star formation triggered by CCC occurred sequentially along this collision path. 
Given the peak column density of $10^{22.6}$ cm$^{-2}$ in N65, the CCC is expected to trigger about 3 to 5 O/B-type stars based on the empirical fitting formula (i.e., ${\rm log(N_{OB})=0.73\ log(N_{H_2})-16.1}$) in \cite{Enokiya2021}, which is consistent with the five signs of massive star formation observed in N65. 
This suggests that the collision of N65a and N65b is consistent with the typical CCC model \citep{Habe1992}, which is undergoing Phase 3 (H\&O Model) after the collision. 
Similar collisions such as M 20 \citep{Torii2011}, RCW 79 \citep{Ohama2018}, RCW 120 \citep{Torii2015} and S235 \citep{Chen2024}, etc., also formed a single or several O/B-type stars.

Figure \ref{fig:trend}(a) presents the distribution of the MST groups along the collision path in terms of the number of group members (N$_{\rm YSO}$) and the ratio of number of Class I to Class II sources (I/II). 
The MST groups are sorted by their galactic longitudes, separating the MST groups into two parts, where M8, M7, M6, and M5 are associated with the N65a molecular cloud, and M1, M2, M3 and M4 are associated with the N65b molecular cloud, with the collision vector pointing from N65a to N65b. 
In the figure, the number of members of the MST group peaks near the collision interface (M1), indicating that CCC effectively triggers star formation at this location. 
The I/II ratio represents the age of the MST group. 
As shown in the figure, all of the young MST groups (M2, M3 and M4, with the I/II ratio of 54\%, 40\% and 30\%, respectively) are distributed in the pre-frontal edge of the N65b molecular cloud, indicating that these star formations were triggered by the CCC very recently.

This conclusion is reinforced by the distribution of spectral indices of the selected YSOs along the galactic longitude. 
We calculated the spectral indices $\alpha$ for the YSOs with K band and IRAC four bands photometric (and MIPS 24 $\mu$m band if possible)  by the following relation (\citealt{Greene1994}), 
\begin{equation}
    \alpha=\frac{d\ {\rm log}(\lambda F_{\lambda})}{d\ {\rm log}\lambda}, 
\end{equation}
where $\lambda$ are the wavelengths in a range of $2.2-8.0$ $\mu$m (or $2.2-24.0$ $\mu$m if band [24] is available), and $F_{\lambda}$ is the flux density at the corresponding wavelength. 
The $\alpha$ indices are obtained by calculating the slope of the least-squares fitting of all available data. 
According to \cite{Greene1994}, they classified the sources with $\alpha>0.3$ as Class I, $0.3>\alpha\geq-0.3$ as Flat-spectrum, $-0.3>\alpha\geq-1.6$ as Class II, and $\alpha<-1.6$ as Class III YSOs. 
As a result, a total of 214 YSOs ($\sim60\%$) have K band and IRAC four bands photometric simultaneously, with $\alpha$ indices spanning the range of $-1.8\leq\alpha\leq3$. 
Figure \ref{fig:trend}(b) shows the distribution of $\alpha$ indices along the galactic longitude. 
The $\alpha$ indices show a distinct gradient change at the collision interface. 
In the N65a-dominated region, the average $\alpha$ index is about $-0.62$ (Class II), indicating that YSOs in this region are generally older (approximately 2 Myr). 
On the other hand, in the N65b-dominated region, the average $\alpha$ index is about $-0.22$ (Flat), indicating that the region has an equal share of young and old YSOs. 
This age-stratified distribution can be explained by the sequential triggered star formation along the collision path, i.e., the YSOs at the post-frontal edge are older because they were triggered to form earlier, while the YSOs at the pre-frontal edge are younger because they are being triggered.

\subsection{CCC scenario in N65}

\begin{figure}[hb]
    \centering
    \includegraphics[width=1\textwidth]{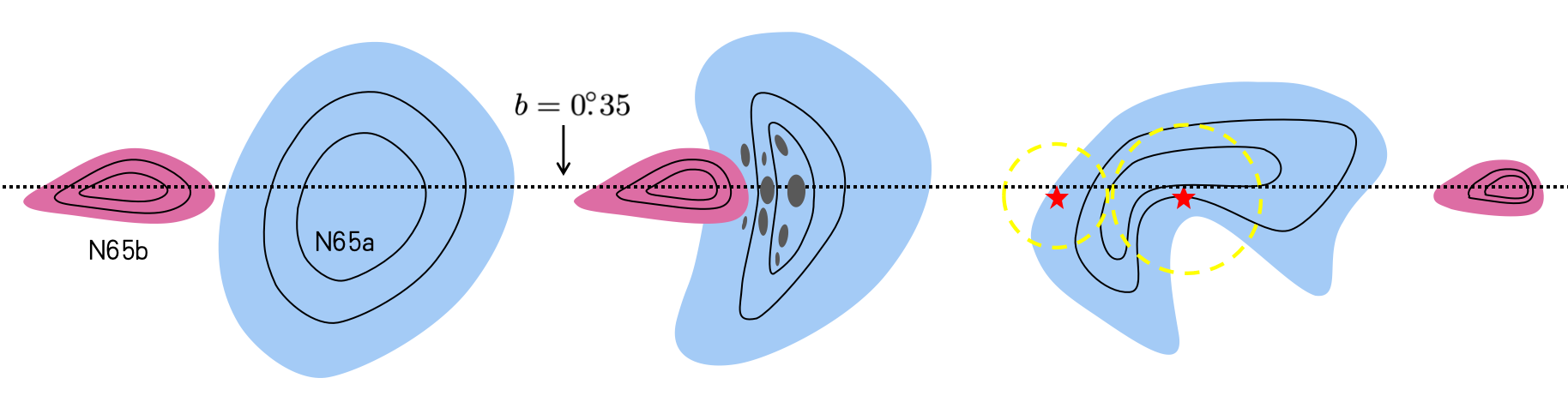}
    \caption{The CCC scenario of the formation and evolution of N65. As shown in phase 1, N65b, as the Bullet Nebula, collides with N65a along the axis of $b=0\fdg35$ in a relative velocity of 8.5 to 12 \kms. In phase 2, the collision preferentially compresses the gas along the axis, creating a dense layer at the colliding interface. Finally, in phase 3, two massive stars with \HII regions are formed along the axis, and N65b separates from N65a due to gravitationally unbound. }
    \label{fig:sketch}
\end{figure}

As mentioned above, the present work reveals a typical CCC process toward two coherent clouds (N65a and N65b) with complementary distribution. 
\cite{Habe1992} presented a simple scenario of CCC: a small cloud collides with a large cloud at supersonic speed, creating a U-shape cavity in the large cloud that compresses the gas to form a dense layer, which in turn collapses to form stars in a wide range of mass. 
Based on this picture of CCC, we present a simplified sketch for the evolution of the collision process of the two clouds in Figure \ref{fig:sketch}. 
As shown in the figure, N65b started to collide with N65a about 2 Myr ago along the galactic latitude of $b=0\fdg35$. 
During the collision, the gas along the collision axis will be compressed first, allowing the dense layer on the axis to accumulate enough gas to form massive stars. 
In the final (present) phase of the collision, the collision-triggered stars (YSOs) are left behind and scattered along the collision path, with only two massive stars remaining on the collision axis and eroding their parent molecular cloud by evolving into \HII regions.

\section{Conclusions} \label{sec:conclusions}

Our CO kinematic analysis, together with the YSO identification, reveals the CCC process of the two clouds, N65a and N65b, and their consequential impact on star formation. 
The important conclusions of this work are summarized as follows: 

\begin{enumerate}
    \item The distinct signatures, such as the bridge feature, the U-shape cavity and the spatial complementary distribution with displacement between N65a and N65b, suggest that they are undergoing CCC process. 
    \item N65a and N65b are gravitationally bound as individual clouds, suggesting that they will collapse to form massive stars. However, they are gravitationally unbound when considered as a whole, indicating that they will separate from each other after the collision. 
    \item The collision between N65a and N65b occurred along the galactic latitude, with the collision axis of $b=0\fdg35$. The collision timescale is estimated to be 1.15 to 2.0 Myr, which is consistent with the dynamical ages of the \HII regions in N65a (0.73 Myr for N65bis and 1.19 Myr for N65, respectively), indicating the CCC-related origin of them. 
    \item A total of 354 YSOs are found in our study region, which are clustered into eight MST groups. They are distributed either over the gas with high velocity dispersion (M1, M3 and M6) or scatter along the collision path (M5, M6, M7, and M8 behind the collision vector, and M1 at the post-frontal edge and M2, M3, and M4 at the pre-frontal edge of N65b), indicating that they are possibly triggered by the CCC. The number of YSOs peaks at the collision interface, and the MST groups close to the pre-frontal edge of N65b are younger. 

    \item Based on the observational evidence above, we can deduce the dynamic interactions and subsequent evolution of the CCC process within the N65 system as follows: initially, N65b and N65a experienced a collision approximately 2 million years ago, occurring along the axis defined by $b=0\fdg35$. This collision subsequently caused a preferential compression of the gas along the collision axis, which in turn led to the formation of two massive stars. Ultimately, the massive stars evolve into the \HII regions, and N65b separates from N65a due to gravitationally unbound.

\end{enumerate}

\begin{acknowledgments}
We acknowledge supports by the National Key R\&D program of China (2022YFA1603102), the National Natural Science Foundation of China (NSFC, grant No. 12473024). X.C. thanks to Guangdong Province Universities and Colleges Pearl River Scholar Funded Scheme (2019). 
This work made use of the data from the Milky Way Imaging Scroll Painting (MWISP) project conducted by Purple Mountain Observatory (PMO). 
MWISP was sponsored by National Key R\&D Program of China with grants 2023YFA1608000 \& 2017YFA0402701 and by CAS Key Research Program of Frontier Sciences with grant QYZDJ-SSW-SLH047. 
This paper made use of the NIR data products from the UKIRT Infrared Deep Sky Survey (UKIDSS) and the archived MIR data obtained with the $Spitzer$ Space Telescope (operated by the Jet Propulsion Laboratory, California Institute of Technology under a contract with NASA). 
The paper made use of the FIR $Herschel$ images at 70, 160, 250, 350, and 500 $\mu$m from the $Herschel$ infrared Galactic Plane Survey (Hi-GAL). $Herschel$ is an ESA space observatory with science instruments provided by European-led Principal Investigator consortia and with important participation from NASA. 
The paper made use of the 1.1 mm dust continuum data from the Bolocam Galactic Plane Survey (BGPS), and the 21 cm radio continuum data from the VLA Galactic Plane Survey (VGPS). 
The paper also made use of the 6.7 GHz methanol maser data from the online tool $MaserDB$. 
\end{acknowledgments}

\bibliography{library}{}
\bibliographystyle{aasjournal}



\end{document}